\documentclass[twocolumn,superscriptaddress,aps,prb]{revtex4-1} 
\usepackage{bm}
\usepackage{amssymb,amsmath}
\usepackage{comment}    
\usepackage{times}
\usepackage{graphicx}
\usepackage{array}

\begin{document}

\title{Thouless conductances of a three-dimensional quantum Hall system}

\author{Chao Zheng}
\email{zhengchaonju@gmail.com}
\affiliation{Zhejiang Institute of Modern Physics, Zhejiang University, Hangzhou 310027, China}

\author{Kun Yang}
\affiliation{Physics Department and National High Magnetic Field Laboratory, Florida State University, Tallahassee, FL 32306, USA}

\author{Xin Wan}
\affiliation{Zhejiang Institute of Modern Physics, Zhejiang University, Hangzhou 310027, China}
\affiliation{CAS Center for Excellence in Topological Quantum Computation, University of Chinese Academy of Sciences, Beijing 100190, China}

\begin{abstract}
We investigate the longitudinal conductance of a disordered three-dimensional (3D) quantum Hall system within a tight-binding lattice model using numerical Thouless conductance calculations.
For the bulk, we confirm that the mobility edges are independent of the propagating directions in this anisotropic system.
As disorder increases, the conductance peak of the lowest subband in the horizontal direction floats to the central subband as in the two-dimensional (2D) case, while there is no clear evidence of floating in the vertical direction. 
We thus conclude that for extended states, the longitudinal conductance in the vertical direction behaves like a quasi-one-dimensional (1D) normal metal, while the longitudinal conductance in the horizontal direction is controlled by layered conducting states stacked coherently.
Inside the quantum Hall gap, we study the novel 2D chiral surface states at the sidewalls of the sample. 
We demonstrate the crossover of the surface states between the quasi-1D metal and insulator regimes, which can be achieved by modifying the interlayer hopping strength and the disorder strength in the model.
The typical behaviors of the Thouless conductance and the wave functions of the surface states in these two regimes are investigated.
Finally, in order to predict the regime of the surface states for arbitrary parameters, we determine an explicit relationship between the localization length of surface states and the microscopic parameters of the model.
\end{abstract}

\date{\today}

\maketitle

\section{Introduction}
\label{sec:intro}
The quantum Hall effect (QHE) in two-dimensional (2D) electron systems originates from discrete Landau levels forming under a strong perpendicular magnetic field~\cite{Das97,Prange12}.
While the physical picture underlying QHE seems specific to 2D, the generalizations of the QHE to 3D systems have also been considered~\cite{Halperin87,Stormer86,Chalker95,Balents96}.
The most straightforward way to construct a 3D QHE is to stack 2D QHE layers along the magnetic field ($z$ axis). As long as the coupling strength is small compared to the 2D quantum Hall gap, we expect the QHE still exists~\cite{Stormer86,Chalker95}.
There are also other schemes to realize a 3D QHE, such as by virtue of formation of spontaneous charge (spin) density wave~\cite{Halperin87} or applying strong magnetic field to topological semimetals~\cite{Wang17,Zhang2017,Uchida2017,Schumann18,Zhang19}.
In the present work, we focus on the first.
The distinct feature of a 2D quantum Hall system is its chiral edge states, which are immune to disorder.
In the 3D case, the chiral edge state of each layer is coupled to neighboring edge states, forming a 2D chiral surface state~\cite{Balents96}.

Experimentally, the 3D QHE was first realized by St\"{o}rmer \textit{et al.}~\cite{Stormer86} in an engineered multilayer quantum well system.
The existence of the 2D chiral surface state in this system was further confirmed in Ref.~\cite{Druist98}.
For real materials, anisotropic layered 3D materials are the most promising candidates to host the 3D QHE.
Signatures of 3D QHE have been found in Bechgaard salts~\cite{Cooper89,Hannahs89}, $\eta$-$\text{Mo}_{\text{4}}\text{O}_{\text{11}}$~\cite{Hill98}, graphite~\cite{Kopelevich03,Bernevig07}, $n$-doped $\text{Bi}_\text{2}\text{Se}_\text{3}$~\cite{Cao12}, $\text{EuMnBi}_\text{2}$~\cite{Masuda16}, and most recently, $\text{ZrTe}_\text{5}$~\cite{Tang19} and $\text{BaMnSb}_\text{2}$~\cite{Liu19}.
Remarkably, the last one provides the first observation of the 2D chiral surface states in a real material.
 These materials offer us great opportunities to study the 3D QHE and its novel surface states.

Up to now, the theoretical understanding of 3D quantum Hall systems mainly comes from a Chalker-Coddington network model~\cite{Chalker95,Kim96,Wang97,Gruzberg97-1,Gruzberg97-2,Cho97,Plerou98} and a continuum model~\cite{Balents96,Mathur97,Balents97,Meir98,Betouras00,Tomlinson05-1,Tomlinson05-2}.
Chalker \textit{et al.}~\cite{Chalker95} investigated the bulk of a 3D quantum Hall system using a generalized 3D network model.
They found that different from the 2D case, the 3D quantum Hall system supports a finite energy range of extended states.
They also introduced a 2D directed network model to describe the chiral surface states.
This model has been extensively studied both analytically~\cite{Gruzberg97-1,Gruzberg97-2} and numerically~\cite{Cho97,Plerou98} later.
Interestingly, it turns out that parallel to the magnetic field in a mesoscopic sample, there exist three distinct regimes of transport, namely, 2D chiral metal, quasi-1D metal, and quasi-1D insulator~\cite{Mathur97,Balents97,Gruzberg97-1,Gruzberg97-2,Cho97,Plerou98}.
Gruzberg, Read, and Sachdev analytically obtained the universal crossover functions of the conductance and its variance between different regimes~\cite{Gruzberg97-1,Gruzberg97-2}.
Some of their results are verified numerically in Refs.~\cite{Cho97,Plerou98}.

The network model is based on the semiclassical picture of electrons in a strong magnetic field and smooth disorder potential~\cite{Chalker88,Kramer05}.
It is, however, only valid when the disorder strength is much smaller than the Landau level spacing~\cite{Chalker95}.
An alternative way to describe the 3D quantum Hall system is using a tight-binding lattice model.
So far, however, there have been very few investigations of such model in the literature~\cite{Wang99}.
Wang \textit{et al.}~\cite{Wang99} studied a tight-binding model by stacking the 2D Hofstadter model along the $z$ direction while keeping interlayer hoppings small.
A phase diagram of the bulk was obtained using the scaling of the Lyapunov exponents in the horizontal direction.
However, we mention that the 3D quantum Hall system is highly anisotropic.
Even without a magnetic field, it has been controversial whether the metal-insulator transition depends on the propagating directions~\cite{Rojo93,Abrikosov94,Zambetaki96}.
It remains unclear whether the mobility edges are the same along the horizontal and vertical directions when a magnetic field is turned on.
Thus, investigations of the model along both directions are indeed necessary.
Most importantly, the chiral surface state has not been studied in Ref.~\cite{Wang99}.

Motivated by recent experimental developments in this direction, in this work, we investigate the longitudinal conductance of a disordered 3D quantum Hall system within a tight-binding model using numerical Thouless conductance calculations.
For the bulk, we emphasize the different behaviors of conductance in the horizontal and vertical directions for this anisotropic system, which is largely ignored in the previous studies~\cite{Chalker95,Wang99}.
The calculated mobility edges along the two directions turn out to be the same under various disorder strengths. 
Their positions are consistent with the results in Ref.~\cite{Wang99}.
As disorder increases, the conductance peak of the lowest subband in the horizontal direction floats to the central subband as in the 2D case, while there is no clear evidence of floating in the vertical direction. 
We thus conclude that for extended states, the longitudinal conductance in the vertical direction behaves like a quasi-1D normal metal, while the longitudinal conductance in the horizontal direction is controlled by layered conducting states stacked coherently.

The 2D chiral surface state has not been studied in the tight-binding model before.
Compared with the 2D directed network model~\cite{Cho97,Plerou98}, studying the surface states within a 3D lattice model is more computationally expensive.
However, the tight-binding model has the advantages that it deals with wave functions directly and its parameters are more experimentally meaningful~\cite{Cho97}.
In the network model, all the information is contained in a single parameter, the transmission coefficient at the saddle point~\cite{Kramer05}.
This is an effective but less direct way to describe the system.
We thus believe our study of the tight-binding model should be useful in guiding experimental efforts of detecting and controlling the chiral surface states.
We demonstrate the crossover of the surface states between the quasi-1D metal and insulator regimes, which can be achieved by modifying the interlayer hopping strength and the disorder strength in the model.
The typical behaviors of the Thouless conductance and the wave functions of the surface states in these two regimes are investigated.
Finally, in order to predict the regime of the surface states for arbitrary parameters, we determine an explicit relationship between the localization length of surface states and the microscopic parameters of the model. 

The rest of the paper is organized as follows. In Sec.~\ref{sec:model} we describe the tight-binding Hamiltonian for 3D quantum Hall system and introduce the numerical Thouless conductance method. In Sec.~\ref{sec:Results} we present our numerical results. The paper is summarized in Sec.~\ref{sec:conclusions}.

\section{Model and Methods}
\label{sec:model}

\subsection{3D lattice model}
\label{subsec:lattice_model}
We consider an electron on an $L_x \times L_y \times L_z$ cubic lattice in the presence of a magnetic field $B \hat{z}$ with tight-binding Hamiltonian
\begin{equation}
\mathcal{H}=-\sum_{\langle i,j\rangle}\left (t_{ij} e^{i\theta_{ij}}c_i^\dagger c_j+h.c.\right)
+ \sum_i \epsilon_i c_i^\dagger c_i,
\label{eq:lattice_model}
\end{equation}
where we have anisotropic nearest-neighboring hopping
\begin{equation*}
t_{ij} = \left \{
\begin{array}{ll}
1 & {\rm \ }i{\rm \ and\ }j{\rm \ are\ horizontal\ nearest\ neighbors,} \\
t_z & {\rm \ }i{\rm \ and\ }j{\rm \ are\ vertical\ nearest\ neighbors,} \\
0 & {\rm \ }i{\rm \ and\ }j{\rm \ are\ not\ nearest\ neighbors.}
\end{array}
\right .
\end{equation*}
We choose Landau gauge $\vec{A}=(0,Bx,0)$ and define $\theta_{ij}=\frac{e}{\hbar}\int_i^j \vec{A}\cdot \mathrm{d} \vec{l}$.
The magnetic flux $\phi$ per unit cell in a horizontal plane is
\begin{equation}
\frac{\phi}{\phi_0}=\frac{Ba^2}{hc/e}=\frac{1}{2\pi}\sum_{\Box} \theta_{ij},
\end{equation}
where $\phi_0 = hc/e$ is the flux quantum, and the disorder potential $\epsilon_i$
are independent variables with identical uniform distribution on $[-W/2,W/2]$.

In the 2D limit with $L_z = 1$, the clean model has a butterfly-like self-similar energy spectrum, as the flux $\phi$ per unit cell varies~\cite{Hofstadter76}.
When the flux $\phi$ per unit cell is chosen as $\phi_0 / N$ for integer $N$, there are exactly $N$ subbands in the spectrum.
The side subbands have Chern number $+1$ each and can be regarded as broadened Landau levels.
The lowest subband, whose localization length is moderate, is suitable for studying the universal behavior of quantum Hall transitions in the presence of disorder.
For small $t_z$, the subband gaps may not close by the dispersion in $z$ axis, so the Hall conductance in the horizontal direction remains quantized.

\begin{figure}
\centering
\includegraphics[width=8.6cm]{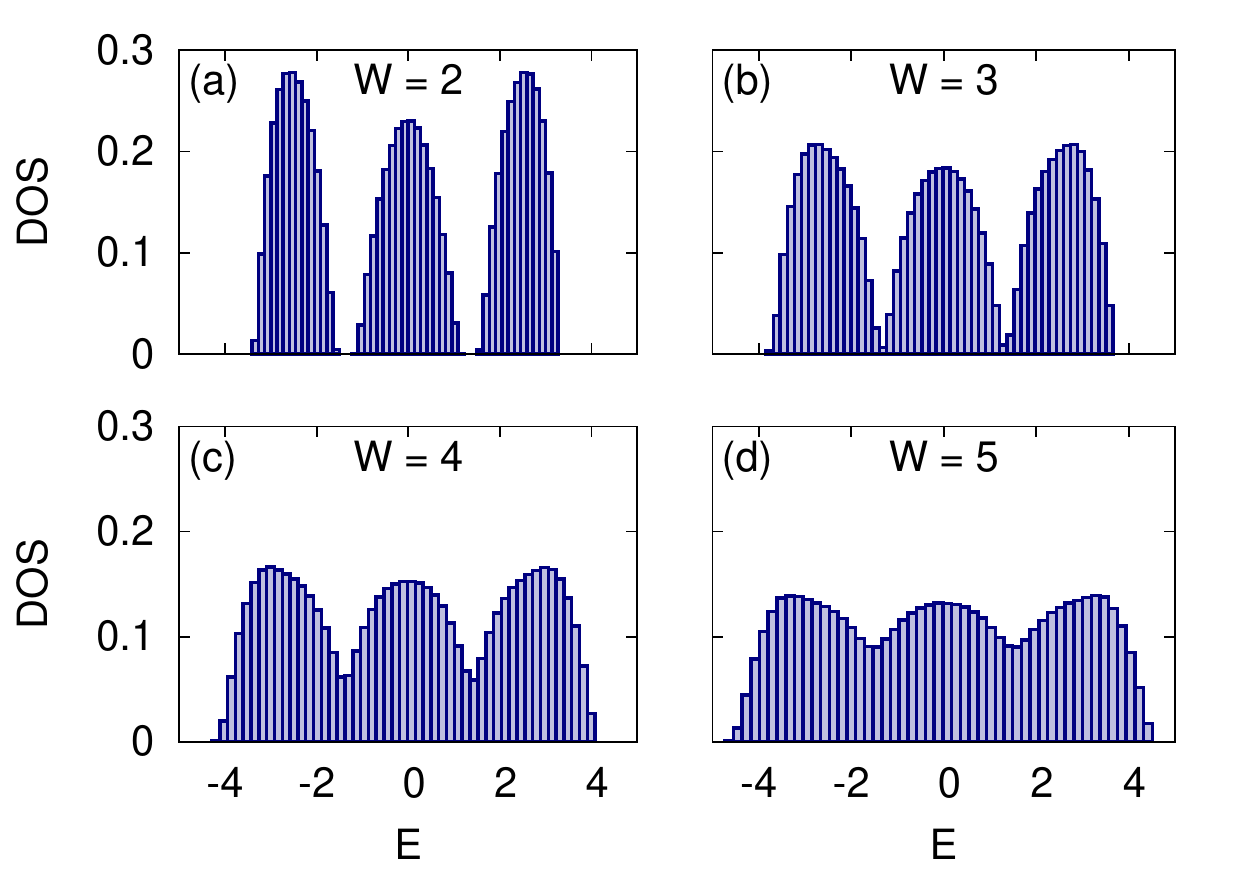}
\caption{Density of states of a $24 \times 24 \times 24$ cubic lattice with $\phi = \phi_0/3$ and $t_z = 0.1$ for (a) $W = 2$, (b) $W = 3$, (c) $W = 4$, and (d) $W = 5$.}
\label{fig:dos3D}
\end{figure}

Figure~\ref{fig:dos3D} shows the evolution of density of states, at $W = 2$-5, for a system with $L = 24$, $\phi = \phi_0/3$, and $t_z = 0.1$.
The gaps between the subbands close at about $W = 3$.
For the small $t_z$, we find the density of states is very similar to that of the corresponding 2D lattice as expected.

\subsection{Thouless Conductance}
\label{sec:Method}

The longitudinal conductance of a finite system can be related to the sensitivity of eigenenergies to changes in the boundary conditions.
The idea came from D. J. Thouless, who argued that the shift in energies of eigenstates due to boundary condition change should be of the order of $\delta E \sim \hbar / \tau$,
where $\tau \sim L^2 / D$ is the time that takes for an electron to
diffuse to the boundary~\cite{Thouless74}.
The diffusion constant $D$ can be related to the longitudinal
conductance by the Einstein relation,
\begin{equation}
\sigma_{xx} = e^2 D \rho,
\end{equation}
where $\rho$ is the density-of-states at the Fermi energy $E_F$.
Therefore, the sensitivity can be quantified in a dimensionless
way~\cite{Licciardello75}
\begin{equation}
\label{thouless}
g_{xx} = {\langle \delta E \rangle \over \langle \Delta E \rangle}
\sim {h \over e^2} \sigma_{xx},
\end{equation}
where $\langle \delta E \rangle$ is the average shift in the energy
level due to a change of boundary conditions along $x$ direction,
and $\langle \Delta E \rangle = 1 / L^d \rho$
the mean energy level separation.
Ando showed that the Thouless number can be applied to 2D
systems in a strong magnetic field, and obtained
\begin{equation}
\label{eq:sigmaxx}
\sigma_{xx} = A {e^2 \over h} g_{xx}, \hspace{1cm} A = {\pi \over 2}
\end{equation}
for peak conductivity for spin-polarized electrons~\cite{Ando83}.
In numerical calculation, the value of $g_{xx}$ depends on the detailed method to evaluate the energy shift.
In general, the proportionality constant $A$ is expected to be of order unity.

As system size increases, the Thouless conductance of a localized state decreases to zero exponentially,
while that of a metallic state does not.
Therefore, the Thouless number can be used to distinguish metallic states from localized states.
Since quantum Hall transition is a special kind of metal-insulator transition, the peak Thouless conductance of a single Landau level is expected to be a universal constant, related to the longitudinal conductivity~\cite{Huo93,Wang96,Wang98,Schweitzer05}.

In this study, we define $\langle \delta E \rangle$ as the arithmetic average of the energy curvature in the boundary-condition space.
Because the time-reversal symmetry is broken in the model, we calculate the curvature in the vicinity of the periodic boundary conductions.
For small systems, the calculation of curvature for random boundary conditions helps suppress the artefacts due to van Hove singularities.

\section{Results}
\label{sec:Results}

\subsection{Evolution of the lowest subband from 2D to 3D}
\label{subsec:2_3D}

We start by stacking $24 \times 24$ square lattices along the magnetic field $\vec{B}$ to a 3D lattice.
In the absence of vertical hopping, the longitudinal conductance perpendicular to $\vec{B}$ scales with the number of layers, while the longitudinal conductance parallel to $\vec{B}$ is zero.
We turn on the vertical hopping to $t_z = 0.1$, still small compared to the horizontal hopping $t = 1$.
To study the crossover from 2D to 3D, we compare the conductance for $L_z = 1$, 2, 3, 4, 6, 8, 12, and 24.
For the 3D system, we calculate Thouless conductance $g_{xx}$ along the $x$-axis by changing the boundary conditions along the $x$ direction and, similarly, $g_{zz}$ by changing the boundary conditions along the $z$ direction.

\begin{figure}
\centering
\includegraphics[width=8.6cm]{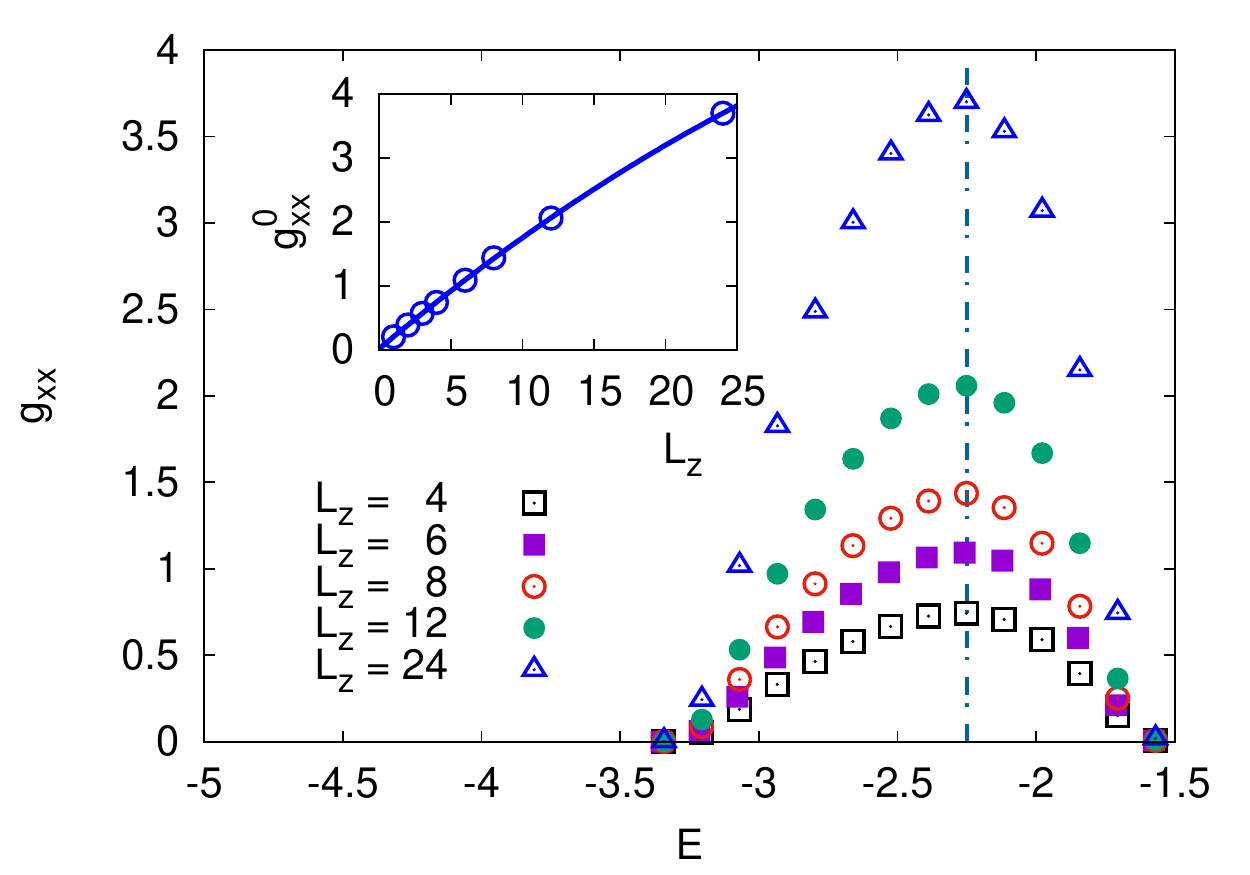}
\caption{Thouless conductance $g_{xx}$, perpendicular to the magnetic field, of the lowest subband in a $24 \times 24 \times L_z$ cubic lattice with $\phi = \phi_0/3$, $W = 2$ and $t_z = 0.1$. The vertical line at $E = -2.25$ indicates the location of the peak of $g_{xx}$. Inset shows the peak conductance $g_{xx}^0$ as a function of $L_z$, which can be fit by $g_{xx}^0(L_z) = 0.020 + 0.188 L_z - 0.00143 L_z^2$ up to the cubic case $L_z = 24$. }
\label{fig:gxxPeak2_3D}
\end{figure}

Figure~\ref{fig:gxxPeak2_3D} plots Thouless conductance along the $x$ direction for the same lowest subband with $\phi = \phi_0/3$ and $W = 2$, for the transverse system size $L_x = L_y = 24$.
The number of states in the lowest subband increases with the number of layers, but the subband width barely changes with the small hopping amplitude in the $z$ direction.
While the shape of $g_{xx}$ is not changed by increasing $L_z$, the peak conductance $g_{xx}^0$ changes.
We find the $g_{xx}^0(L_z)$ can roughly be fit by
\begin{equation}
g_{xx}^0(L_z) = 0.020 + 0.188 L_z - 0.00143 L_z^2,
\end{equation}
for $L_z$ up to 24.
The fit suggests that the contribution from each layer is roughly the same, $g_{xx}^0 = 0.188 \pm 0.001$, with negligible corrections in $L_z$, even for $L_z \sim L_x$.
In addition, $g_{xx}^0$ is also expected to have a weak $L_x$ dependence, which is not important to the present study.

\begin{figure}
\centering
\includegraphics[width=8.6cm]{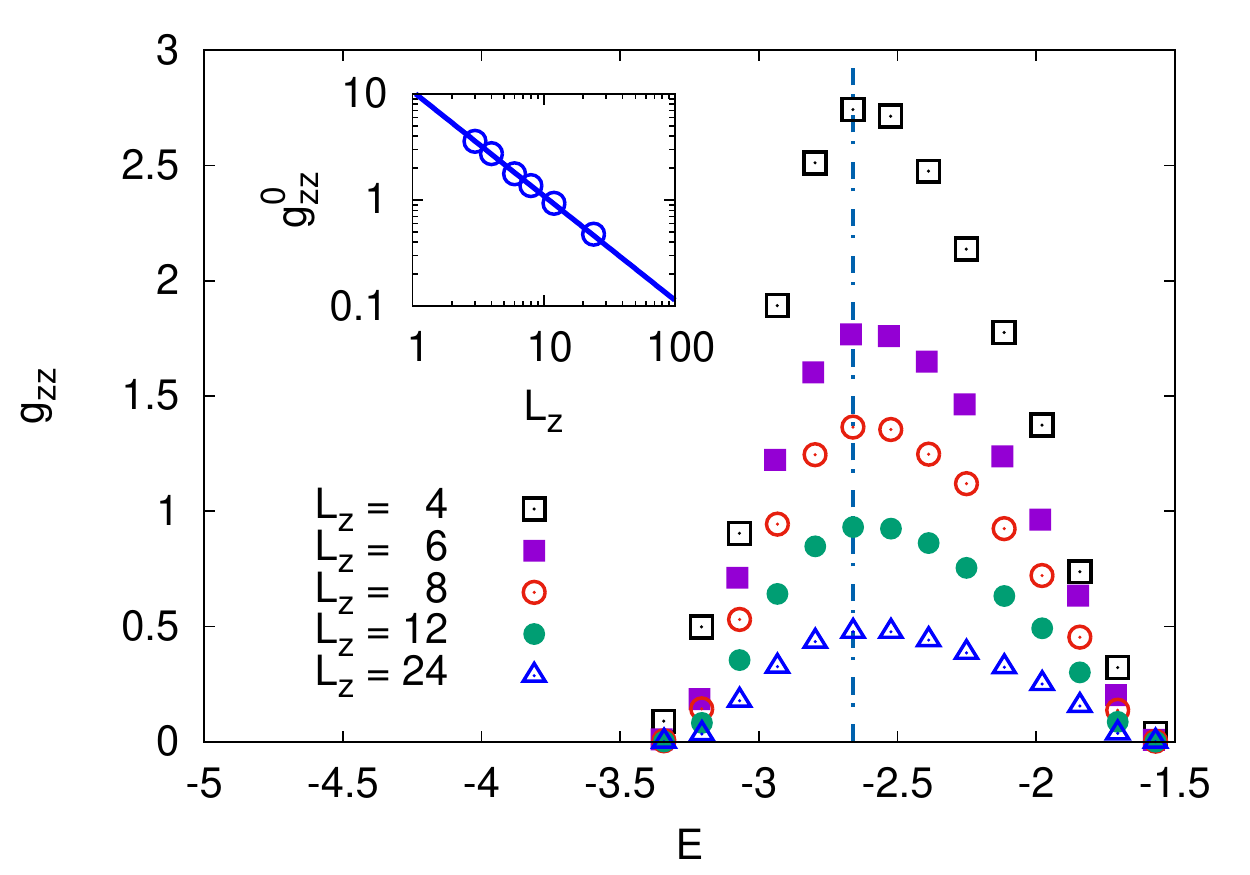}
\caption{Thouless conductance $g_{zz}$, parallel to the magnetic field, of the lowest subband in a $24 \times 24 \times L_z$ cubic lattice with $\phi = \phi_0/3$, $W = 2$ and $t_z = 0.1$.  The vertical line at $E = -2.66$ indicates the location of the peak of $g_{zz}$. Inset shows the peak conductance $g_{zz}^0$ as a function of $L_z$, which can be fit by $g_{zz}^0(L_z) = 10.6 L_z^{-0.983}$.}
\label{fig:gzzPeak2_3D}
\end{figure}

We now turn to Thouless conductance along the $z$ direction, along which the magnetic field aligns and the hopping is weak.
The results are shown in Fig.~\ref{fig:gzzPeak2_3D}.
Like $g_{xx}$, the vertical Thouless conductance develops a hump, whose shape remains unchanged as $L_z$ increases.
But unlike $g_{xx}$, the peak conductance $g_{zz}^0$ decreases with $L_z$, and our power-law fit to
\begin{equation}
g_{zz}^0(L_z) = g_0 L_z^{-\alpha},
\end{equation}
gives $\alpha = 0.983 \pm 0.015$, which suggests that $g_{zz}^0$ is inversely proportional to $L_z$ within error bars.

We note that the peak of the lowest band in the density of states is located around $E_{p} = -2.66$, which coincides with the peak of $g_{zz}$, but not $g_{xx}$, which peaks around $E_{p}^{(x)} = -2.25$. There is no shift of peak in either $g_{zz}$ or $g_{xx}$, as $L_z$ increases from 1 to 24. At $E_{p} = -2.66$, the peak $g_{zz}$ is inversely proportional to $L_z$, indicating the system is in a metallic phase. The shift, or floating, of the $g_{xx}$ peak is expected to originate from the floating of extended states in the 2D limit~\cite{Khmelnitskii84,Laughlin84,Yang96}, as we will discuss more in the following subsection as disorder increases. It also suggests that the $g_{xx}$ in the multilayer cases is dominated by conducting states in each 2D layer. 
The deviation of $g_{xx}^0$ from linear behavior suggests that the coherence of extended states along the $z$ axis plays a role in the longitudinal conductance in the $x$-$y$ plane.

\begin{figure}
\centering
\includegraphics[width=8.6cm]{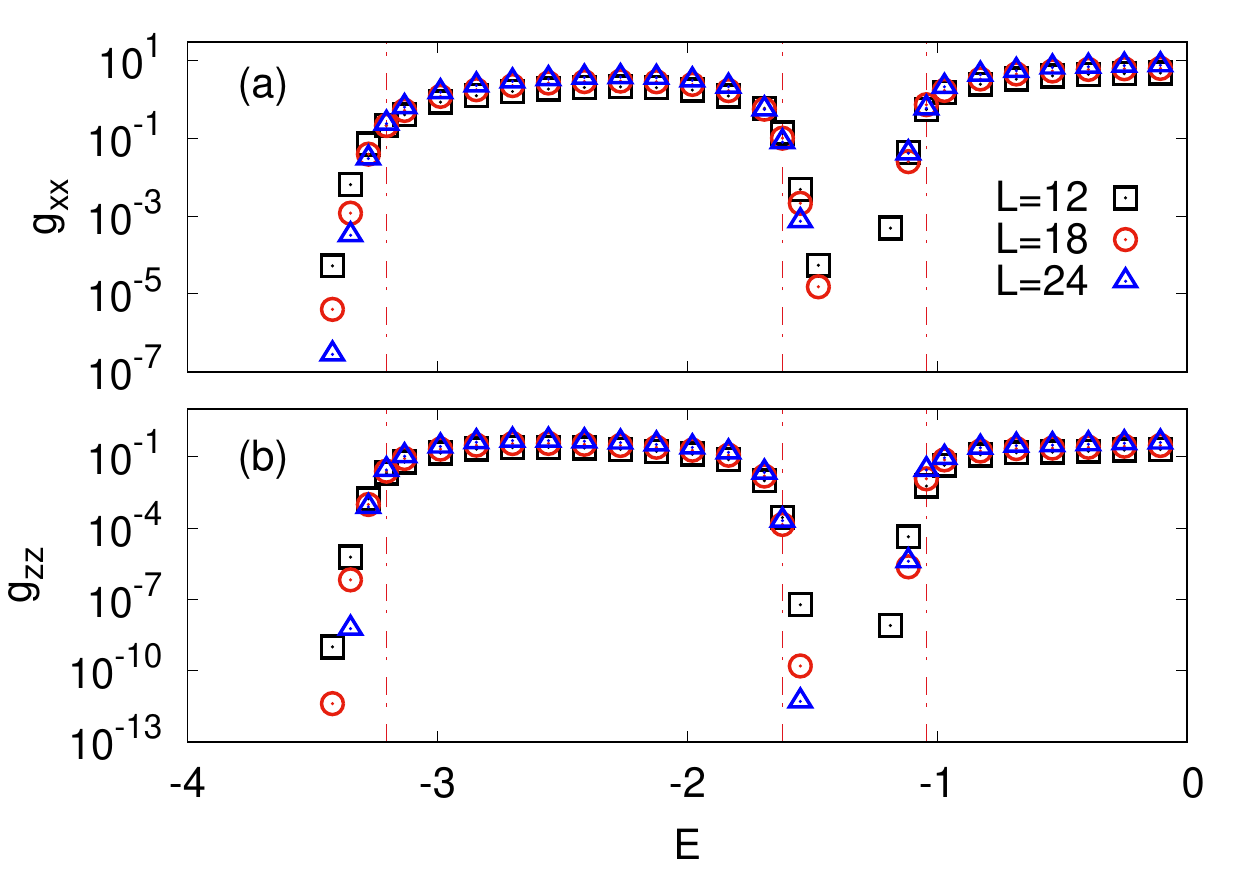}
\caption{Thouless conductance (a) $g_{xx}$, perpendicular to the magnetic field and (b) $g_{zz}$, parallel to the magnetic field, of the lowest subband and a half of the central band in an $L \times L \times L$ cubic lattice with $\phi = \phi_0/3$, $t_z = 0.1$, and $W = 2.0$ for $L = 12$, 18, 24. The vertical lines at $E = -3.20$, $-1.62$, and $-1.05$ indicate the locations of mobility edges, as the size dependence of both $g_{xx}$ and $g_{zz}$ differs on the two sides of the energies. The lines define four different phases along the horizontal axis. The phases from left to right are insulator, metal, quantized Hall insulator, and metal, respectively.}
\label{fig:MITw2.0}
\end{figure}

In the 2D case, there is only one critical energy in the lowest subband, at which we find extended states even in the thermodynamic limit.
When Fermi energy moves across the critical energy, an integer quantum Hall plateau transition occurs.
This, however, is quite different from the 3D case, where a range of metallic states exist~\cite{Chalker95,Wang99}; in fact, we already see that the peaks of $g_{xx}$ and $g_{zz}$ differ.
To estimate the energy range for the metallic phase at $W = 2.0$, we explore the finite-size effect of the conductance for the cubic lattice with $L_x = L_y = L_z = L$.
Figure~\ref{fig:MITw2.0} plots both $g_{xx}$ and $g_{zz}$ as a function of energy for the lower half of the spectrum. We choose, here, $\phi = \phi_0/3$, $t_z = 0.1$, and $W = 2.0$ as before.
The size dependence of the Thouless conductance along two perpendicular directions remains to be the same, indicating the mobility edges are the same in the two directions.
We can clearly identify two insulating regimes and two metallic regimes, separated by $E = -3.20$, $-1.62$, and $-1.05$, as indicated by dot-dashed lines in Fig.~\ref{fig:MITw2.0}.
The mobility edges found here are consistent with the results in Ref.~\cite{Wang99}.
The phase at $-1.62 < E < -1.05$, which includes the band gap, is accompanied by the quantization of Hall conductance and is, therefore, a quantized Hall insulator. We will explore the phase in greater detail in the following subsections.

\subsection{Disorder effects in the bulk}
\label{subsec:3D}

So far, we find that for a relatively small disorder $W = 2.0$, energy states in most of the lowest subband are in a metallic phase. But the phase differs from a normal metallic one. Based on the bulk Thouless conductance, we can argue that the longitudinal conductance along the magnetic field $\vec{B}$ behaves like that of a quasi-1D normal metal. The longitudinal conductance perpendicular to $\vec{B}$ is controlled by layered conducting states stacked coherently.
We now focus on $L \times L \times L$ cubic samples and study the Thouless conductance along $x$ and $z$ directions.
In particular, we are interested in the manifestation of the quantized Hall phase as disorder varies.

\begin{figure}
\centering
\includegraphics[width=8.6cm]{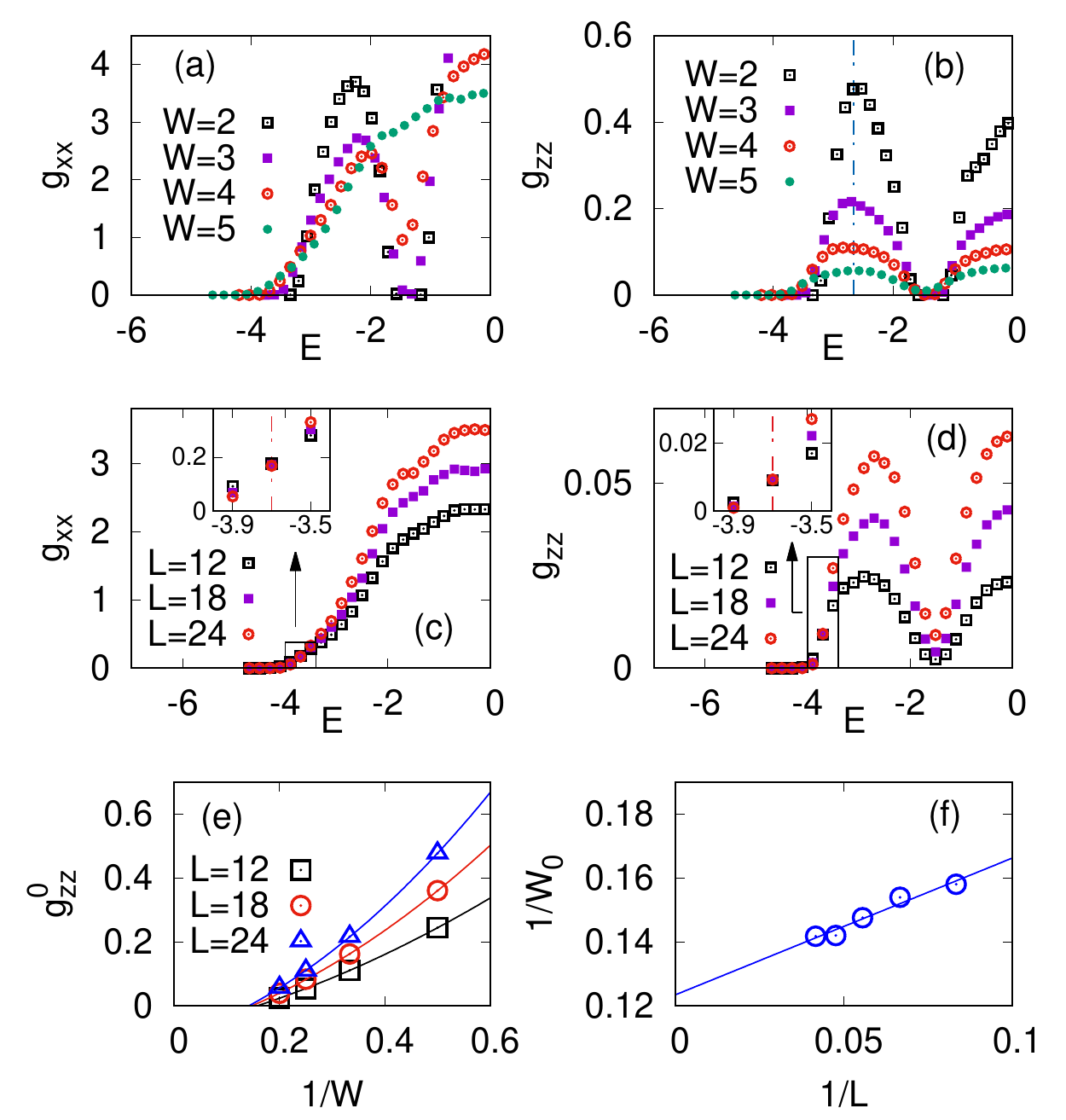}
\caption{
Thouless conductance in an $L \times L \times L$ cubic lattice with $\phi = \phi_0/3$, $t_z = 0.1$.
(a) Thouless conductance $g_{xx}$ and (b) $g_{zz}$ of the lowest subband and a half of the central band with $L = 24$ for $W = 2$-5. 
The dot-dashed line indicates the peak of $g_{zz}^0$ as mentioned in Fig.~\ref{fig:gzzPeak2_3D}.
(c),(d) Finite-size scaling of $g_{xx}$ and $g_{zz}$ at $W=5$, for $L = 12$, 18, 24. 
The insets show enlarged views around the mobility edge $E=-3.7$, the size dependence of both $g_{xx}$ and $g_{zz}$ differs on the two sides of the mobility edge.
(e) Peak conductance $g_{zz}^0$ as a function of $1/W$ for various $L$. We fit the data by quadratic curves with horizontal intercepts $1/W_0(L)$.
(f) The dependence of $1/W_0$ as a function of $1/L$, which can be extrapolated to $L \rightarrow \infty$ at $1/W_0 = 0.123 \pm 0.003$. }
\label{fig:Peak3D}
\end{figure}

Figure~\ref{fig:Peak3D}(a) shows $g_{xx}$ for the lowest subband and part of the central subband
in the case of $\phi = \phi_0/3$ and $L = 24$ for $W = 2$-5.
As disorder increases, the conductance peak of the lowest subband floats to the central subband,
as found in the 2D case~\cite{Yang96}.
No peak can be identified for the lowest subband at $W = 5$, as it merges with the central subband.
In the 3D case we find that, before the merge, the peak value $g_{xx}^0$ increases linearly as $L$ increases from 12 to 24, which is consistent with metallic behavior.

Similarly, we compare $g_{zz}$ for various $W$ with $L = 24$ in Fig.~\ref{fig:Peak3D}(b).
Due to the small hopping strength in the $z$ direction, $g_{zz}$ is much smaller than $g_{xx}$.
The distinct feature of $g_{zz}$ is that the deep dip separating the lowest subband and the central subband persists to $W = 5$, when the two subbands are not resolvable in $g_{xx}$.
This raises the question whether the system is metallic in the $x$ direction while insulating in the $z$ direction near the original band gap when $W=5$.
Therefore, we carry out the finite-size study of $g_{xx}$ and $g_{zz}$ at $W=5$ in Figs.~\ref{fig:Peak3D}(c) and (d).
Near the original band gap at $E=-1.4$, although $g_{zz}$ is small, both $g_{xx}$ and $g_{zz}$ grows when the system size increases.
Thus the system is metallic along both directions.
At $W=5$, the disorder couples the three subbands together to form a large energy band~\cite{Wang99}.
There is only one metal-insulator transition at $E=-3.7$, this value agrees well with the results in Ref.~\cite{Wang99}. 
The mobility edge is the same along the two perpendicular directions, as shown in the insets of Figs.~\ref{fig:Peak3D}(c) and (d).
Combining with Fig.~\ref{fig:MITw2.0}, our results suggest that the mobility edges are the same along the horizontal and vertical directions in this anisotropic system, which is unsettled in the literature~\cite{Chalker95,Wang99}.
We note that it is also interesting to investigate the localization length and the critical exponents along the two perpendicular directions in this system.
This is beyond the scope of present work, and we leave it for future study.

Another feature of $g_{zz}$ is that there is no clear evidence of the floating of the $g_{zz}$ peak of the lowest subband,
in contrast to that of $g_{xx}^0$, as we have commented in the previous subsection.
The peak value $g_{zz}^0$ also scales linearly with $L$, showing metallic behavior
for the range of disorder strength we consider.
Fig.~\ref{fig:Peak3D}(e) plots $g_{zz}^0$ as a function of $1 /W$ for various $L$.
The curves can be fit by quadratic functions, which have size-dependent horizontal intercepts $1/W_0(L)$.
This is an indication of a metal-insulator transition at the band center at sufficiently small $1/W$,
or sufficiently large $W$.
We plot the intercept $1/W_0$ against $1/L$ and fit the data by a straight line, which has a vertical intercept
$1/W_0 = 0.123 \pm 0.003$.
In other words, when disorder strength is greater than $W_c = 8.1 \pm 0.2$,
we expect that the system shows insulating behavior at the peak location of $g_{zz}$,
or $E = -2.66$ as we have obtained earlier.

Note that we have discussed two methods to detect the boundaries of the metallic phase of the lattice model.
In Fig.~\ref{fig:MITw2.0}, Figs.~\ref{fig:Peak3D}(c) and~\ref{fig:Peak3D}(d), we have analyzed the dependence of $g_{xx}$ and $g_{zz}$ on system size,
from which we can extract the mobility edges.
This approach is expected to be more effective at small disorder,
where the phase boundaries in the phase diagram in the $W$-$E$ plane depend weakly on the disorder~\cite{Wang99}.
On the other hand, in Figs.~\ref{fig:Peak3D}(e) and (f), we have first studied the dependence of $g_{zz}$ on
disorder strength for a given system size and extrapolate the characteristic disorder that suppresses
$g_{zz}$ to zero.
We have then carried out the finite-size scaling of the characteristic disorder to
extract the critical disorder in the thermodynamic limit.
This method is expected to work more effectively at large disorder,
where the phase boundary in the phase diagram in the $W$-$E$ plane is almost flat as energy changes~\cite{Wang99}.

\subsection{Chiral surface states}
\label{subsec:surface}

Our main interest so far has been focused on the metallic behavior and the metal-insulator transition of the 3D system.
Another interesting transport property is, however, associated with the surface of a 3D quantum Hall system.
This phase is located at the gap region between the lowest subband and the central subband, characterized by a Chern number 1 for each layer perpendicular to the magnetic field.
With open boundary conditions as in real experimental situations, the chiral edge state of each layer is coupled to neighboring edge states, forming a chiral surface state.

The transport properties of the chiral surface states are highly anisotropic in the presence of disorder~\cite{Balents96}.
Perpendicular to the magnetic field, due to the chiral nature of the edge states, the transport is ballistic with a velocity $v$.
Parallel to the magnetic field, the localization effect is suppressed by the unidirectional transport in the $x$ direction.
In order to make quantum interference happen, an electron has to circumnavigate the sample and return to its starting point.
This can never happen in an infinite sample.
Therefore, it has been argued that for an infinite sample, the transport along the $z$ direction is always diffusive, regardless of the disorder strength~\cite{Balents96,Kim96}.
Using the Einstein relation, we can obtain the conductivity of the 2D sheet in the $z$ direction~\cite{Balents96}
\begin{equation}
  \sigma_{zz}=e^2 \rho D=\frac{D}{v} \frac{e^2}{h}=\sigma \frac{e^2}{h},
\end{equation}
where $D$ is the diffusion constant, $\rho =1/hv$ the density of states, and $\sigma=D/v$ the dimensionless conductivity.

Interestingly, for a mesoscopic sample, depending on its vertical sheet conductivity $\sigma_{zz}$, the circumference $C$, and the height $L_z$, there are three distinct regimes of transport in the $z$ direction connected by universal crossovers, namely, 2D chiral metal, quasi-1D metal, and quasi-1D insulator~\cite{Mathur97,Balents97,Gruzberg97-1,Gruzberg97-2,Cho97,Plerou98}.
The time needed for an electron to circle the sample is $\tau=C/v$.
Within this time, the electron will diffuse a distance
\begin{equation}
\label{eq:L0}
  L_0=\sqrt{D\tau}=\sqrt{DC/v}
\end{equation}
in the $z$ direction.
If $L_z \ll L_0$, then the system is a 2D chiral metal.
In this circumstance, the electron diffuses out of the sample without a complete round-trip of the circumference.
For finite $C$ with very long $L_z$, the system is of quasi-1D nature, we expect the surface state is localized along the $z$ direction. Typically, the localization length $\xi$ of such quasi-1D system is proportional to its 1D conductivity, which can be written as
\begin{equation}
\label{eq:xi}
  \xi=2C\sigma=2DC/v.
\end{equation}
For $L_0 \ll L_z \ll \xi$, the system is a quasi-1D conductor. In such condition, the electron circles around the sample many times before diffusing out. For $L_z \gg \xi$, the interference effect becomes dominant, finally turns the system into a quasi-1D insulator.

Numerically, the chiral surface state in a mesoscopic system was previously studied by the 2D directed network model in Refs.~\cite{Cho97,Plerou98}. 
It has not been studied in the tight-binding model before.
Compared with the network model, the tight-binding model has the advantages that it deals with the wave functions directly and its parameters are more experimentally meaningful~\cite{Kramer05,Cho97}.
For our lattice model, the diffusion constant $D$ of the surface state in the $z$ direction depends on the interlayer hopping strength $t_z$, the disorder strength $W$, and the ballistic velocity $v$ of the edge states.
By adopting the result derived in a 2D continuum model~\cite{Balents96,Tomlinson05-1}, we have
\begin{equation}
\label{eq:D}
  D\varpropto\frac{t_z^2}{W^2}v.
\end{equation}
Substituting Eq.~(\ref{eq:D}) into Eq.~(\ref{eq:L0}) and (\ref{eq:xi}), we obtain
\begin{equation}
\label{eq:L0_2}
  L_0\varpropto\sqrt{\frac{t_z^2}{W^2}C}
\end{equation}
and
\begin{equation}
\label{eq:xi_2}
  \xi\varpropto\frac{t_z^2}{W^2}C.
\end{equation}
Thus both $L_0$ and $\xi$ only depend on $t_z$, $W$ and $C$, they are independent of $v$.

In the following, we demonstrate the metal-insulator crossover between the quasi-1D metal and insulator regimes of the surface states, by fixing the dimensions of the system, only changing $t_z$ and $W$.
We note that to enter into the 2D chiral metal regime, one requires $L_z \ll L_0$.
To increase $L_0$, we can simply increase $t_z$ and decrease $W$. But this also tends to drive the system into the ballistic regime, whereas the 2D chiral metal regime is of diffusive nature.
Another way to increase $L_0$ is to increase $C$. However, due to the square root in Eq.~(\ref{eq:L0_2}), the system size needed is numerically quite challenging for the Thouless conductance calculation~\cite{Cho97,Plerou98}.
Therefore, we focus on the quasi-1D metal and insulator regimes in the following sections.

\subsubsection{Quasi-1D metal}
\label{subsubsec:1D metal}

\begin{figure}
\centering
\includegraphics[width=8.6cm]{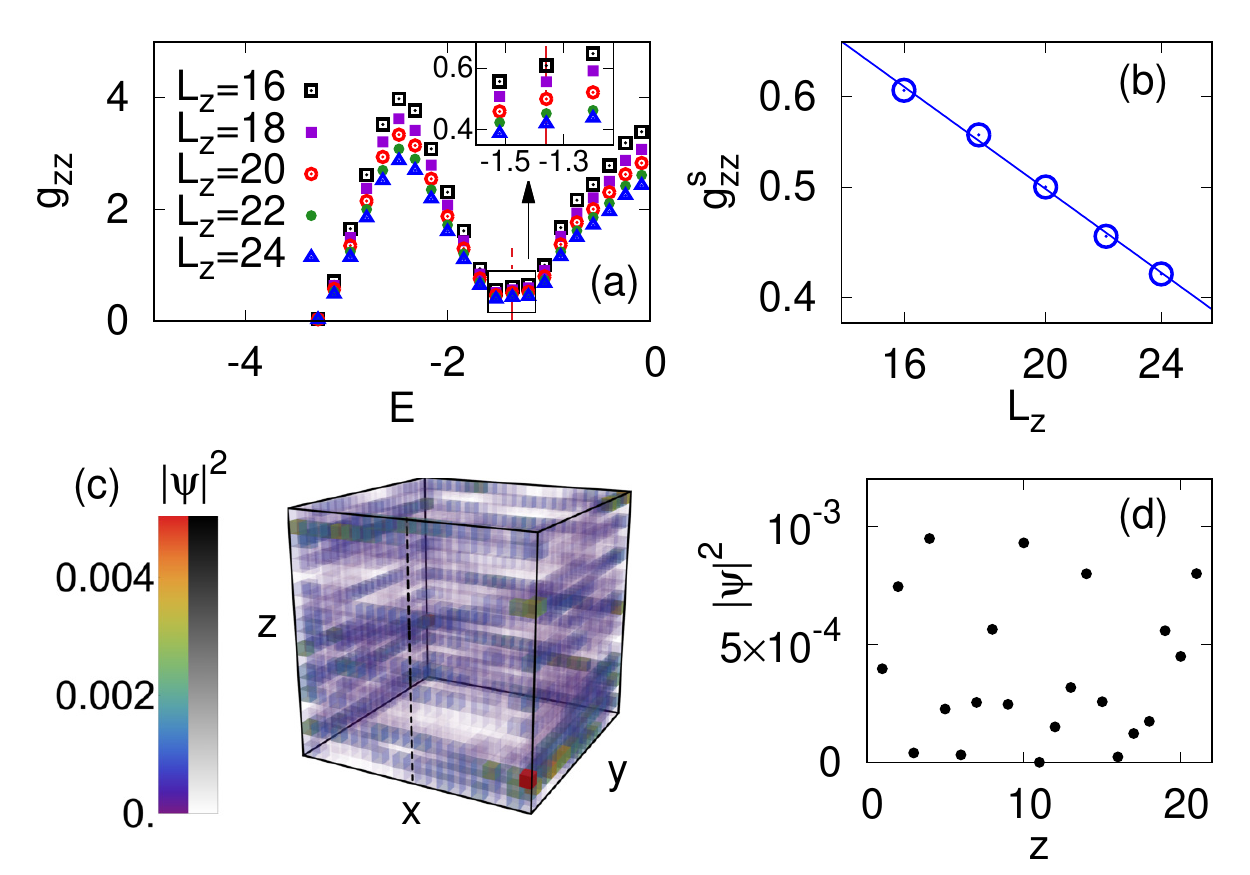}
\caption{
Thouless conductance $g_{zz}$ and wave function of the surface state in the quasi-1D metallic regime with $\phi = \phi_0/3$, $t_z = 0.2$ and $W = 1.3$.
(a) Thouless conductance $g_{zz}$ of the lowest subband and a half of the central band in a $21 \times 21 \times L_z$ cubic lattice with open boundary conditions along $x$ and $y$ directions.
Inset shows a zoom-in view around the plateau region inside the spectral gap.
(b) Log-log plot of the plateau value $g^s_{zz}$ at $E = -1.36$ [indicated as the vertical dot-dashed line in (a)] as a function of $L_z$, which can be fit by $g_{zz}^s(L_z) = 7.97 L_z^{-0.926}$.
(c) The probability density $\left| \psi \right|^2$ for a particular disorder realization in a $21 \times 21 \times 21$ cubic lattice at $E = -1.36$.
The result is obtained by exact diagonalization under open boundary conditions in the $x$ and $y$ directions and periodic boundary condition in the $z$ direction.
Each lattice point is represented by a small cube, whose color and opacity depends on the value of $\left| \psi_i \right|^2$.
The color and opacity bar is given on the left of the plot.
(d) The probability density $\left| \psi \right|^2$ along the dashed vertical line in (c).
}
\label{fig:1D_metal}
\end{figure}

We first present the results in the quasi-1D metal regime.
To study the surface states, we apply open boundary conditions to the $x$ and $y$ directions and calculate Thouless conductance $g_{zz}$ in the $z$ direction.
Figure~\ref{fig:1D_metal}(a) shows $g_{zz}$ of the lowest subband and a half of the central band in a $21 \times 21 \times L_z$ cubic lattice with $\phi = \phi_0/3$, $t_z = 0.2$, and $W = 1.3$.
$g_{zz}$ develops plateaus inside the spectral gaps of the system, which are the contributions from the surface states.
We show an enlarged view of this region in the inset of Fig.~\ref{fig:1D_metal}(a).
Figure~\ref{fig:1D_metal}(b) plots the plateau value $g^s_{zz}$ at $E = -1.36$ as a function of system height $L_z$.
 The log-log plot suggests a power-law decrease of $g^s_{zz}$ with respect to $L_z$, which can be fit to
\begin{equation}
  g_{zz}^s(L_z) = g_0 L_z^{-\alpha},
\end{equation}
with $\alpha=0.926 \pm 0.031$.
This "ohmic" dependence of conductance suggests that the system is a conductor in the $z$ direction.

The transport behavior is determined by the localization properties of the wave function.
In Figs.~\ref{fig:1D_metal}(c) and ~\ref{fig:1D_metal}(d), we plot the wave function in a $21 \times 21 \times 21$ cubic lattice at $E = -1.36$ for a specific disorder configuration.
 One can find that the wave function is extended over the whole sheath of the sample, indicating its conducting nature.

\subsubsection{Quasi-1D insulator}
\label{subsubsec:1D insulator}

\begin{figure}
\centering
\includegraphics[width=8.6cm]{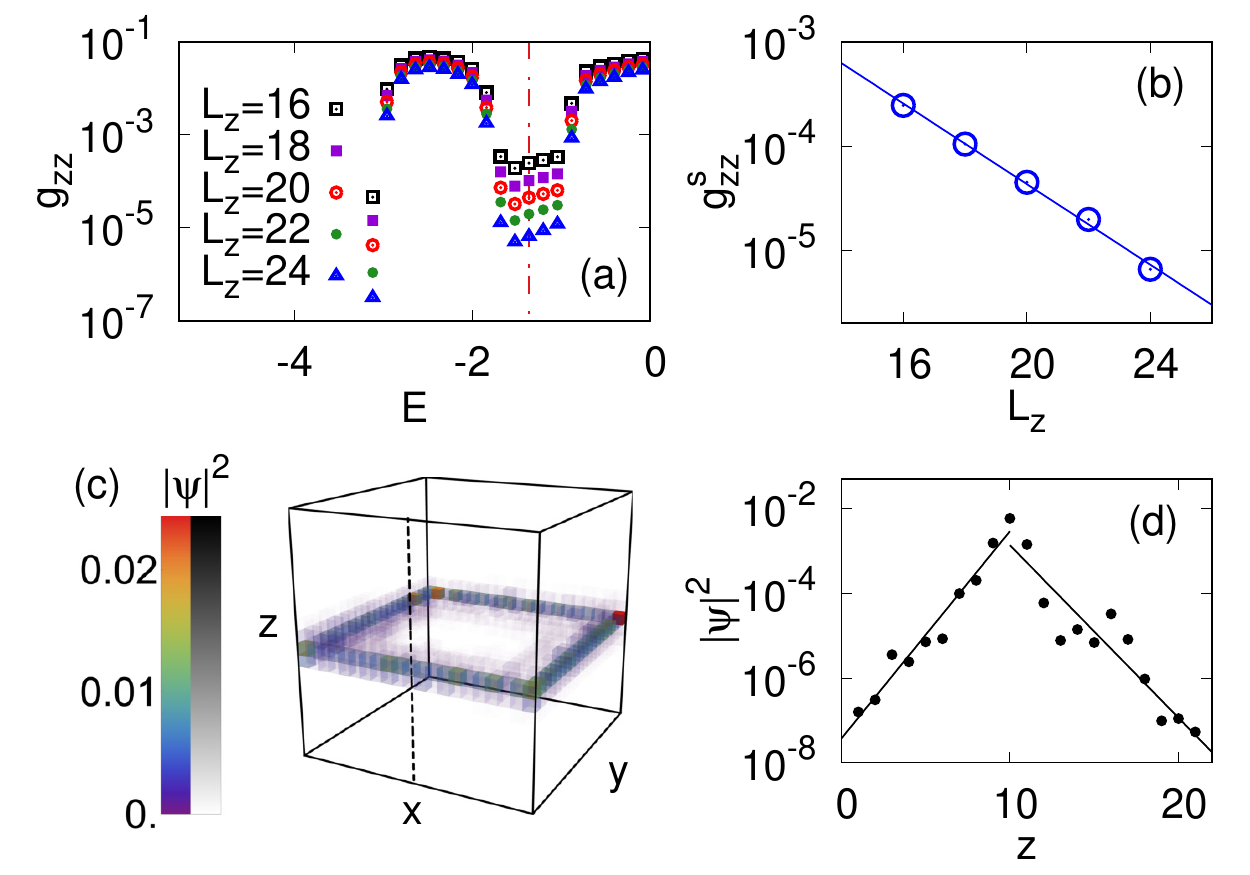}
\caption{
Thouless conductance $g_{zz}$ and wave function of the surface state in the quasi-1D localized regime with $\phi = \phi_0/3$, $t_z = 0.04$ and $W = 2$.
(a) Semi-log plot of Thouless conductance $g_{zz}$ in a $21 \times 21 \times L_z$ cubic lattice with open boundary conditions along $x$ and $y$ directions.
(b) Semi-log plot of the plateau value $g^s_{zz}$ at $E = -1.36$ [indicated as the vertical dot-dashed line in (a)] as a function of $L_z$, which can be fit by $g_{zz}^s(L_z) = 0.320 e^{-L_z/2.24}$.
(c) The probability density $\left| \psi \right|^2$ for a particular disorder realization in a $21 \times 21 \times 21$ cubic lattice at $E = -1.36$, plotted in the same way as Fig.~\ref{fig:1D_metal}(c).
(d) The semi-log plot of probability density $\left| \psi \right|^2$ along the dashed vertical line in (c).
The fit to the wings with an exponential decay $e^{-2z/\xi}$ yields $\xi=1.77 \pm 0.13$ for the left wing and $\xi=2.13 \pm 0.26$ for the right wing.
}
\label{fig:1D_insulator}
\end{figure}

Next, we present the results in the quasi-1D insulator regime.
The quasi-1D insulator regime can be achieved by decreasing the interlayer hopping strength $t_z$ and increasing the disorder strength $W$.
This is demonstrated in Fig.~\ref{fig:1D_insulator}, in which $\phi = \phi_0/3$, $t_z = 0.04$, and $W = 2$.
Figure~\ref{fig:1D_insulator}(a) illustrates the height dependence of $g_{zz}$ in a $21 \times 21 \times L_z$ cubic lattice, note that we use semi-log coordinate in the plot.
Since a quasi-1D system approximately favors a log-normal curvature distribution in the localization regime, here we use geometric averages for $\langle \delta E \rangle$~\cite{Casati94}.
The plateau value $g^s_{zz}$ at $E = -1.36$ as a function of $L_z$ is shown in Fig.~\ref{fig:1D_insulator}(b).
The semi-log plot clearly shows an exponential decay of $g^s_{zz}$ with $L_z$, indicating it is an insulator in the $z$ direction. 
An explicit exponential fit to
\begin{equation}
  g_{zz}^s(L_z) = g_0 e^{-L_z/\xi},
\end{equation}
gives a localization length $\xi=2.24 \pm 0.07$.
We notice that $\xi$ is about $1/10$ of $L_z$,  thus the system is deep in the localization regime in the present case.

In Fig.~\ref{fig:1D_insulator}(c), we show a 3D plot of the wave function at $E = -1.36$ in a $21 \times 21 \times 21$ cubic lattice for a particular disorder realization.
In this case, the surface state is concentrated to a few layers in the $z$ direction.
Figure~\ref{fig:1D_insulator}(d) plots the probability density along the dashed vertical line in Fig.~\ref{fig:1D_insulator}(c).
One can also obtain the localization length from the wave function by using an exponential fit
\begin{equation}
  \left| \psi(z)\right|^2=\left| \psi_0\right|^2 e^{-2\left| z-z_0\right|/\xi},
\end{equation}
where $z_0$ is the maximum position of $\left| \psi \right|^2$.
The fit yields $\xi=1.77 \pm 0.13$ for the left wing and $\xi=2.13 \pm 0.26$ for the right wing of the wave function, which is roughly consistent with the result from the Thouless conductance.

\subsubsection{Determining the relationship between the localization length and the microscopic parameters of the model}
\label{subsubsec:kesi_vs_parameters}

So far, we explore the surface states using two specific sets of parameters which belong to quasi-1D metal and insulator regimes, respectively.
For arbitrary parameters, in order to determine the regime of the system, we need to compare the system height $L_z$ with the localization length $\xi$.
As mentioned above, the localization length depends on the interlayer hopping strength $t_z$, the disorder strength $W$, and the circumference $C$ of the system [see Eq.~(\ref{eq:xi_2})].
We are now in a position to verify Eq.~(\ref{eq:xi_2}) numerically.
In the end, we are able to predict the regime of the system for arbitrary parameters.

\begin{figure}
\centering
\includegraphics[width=8.6cm]{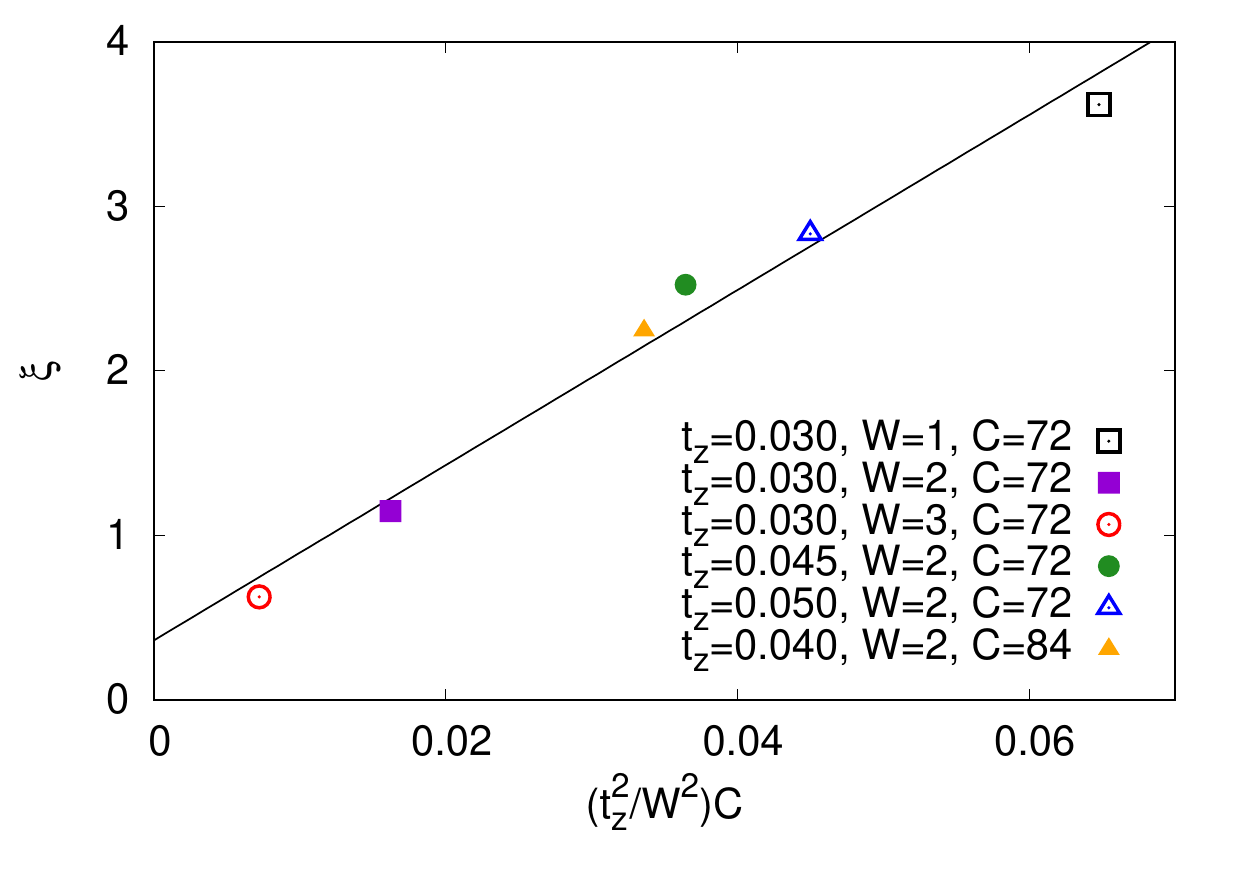}
\caption{
Relationship between the localization length $\xi$ of the surface states and the microscopic parameters of the model. We determine the localization length from the height dependence of Thouless conductance $g_{zz}$ in an $L \times L \times L_z$ cubic lattice, in the same way as Fig.~\ref{fig:1D_insulator}. Here $C$ is the circumference of the sample, which equals $4L$ in the present case. The localization length $\xi$ has a linear dependence on $(t_z^2/W^2)C$.
}
\label{fig:kesi_vs_parameters}
\end{figure}

To do this, we repeat the procedure in Sec.~\ref{subsubsec:1D insulator}, and determine the localization length from the height dependence of Thouless conductance in an $L \times L \times L_z$ cubic lattice.
The sets of parameters we choose and the final result are presented in Fig.~\ref{fig:kesi_vs_parameters}.
The localization length $\xi$ indeed has a linear dependence on $(t_z^2/W^2)C$, which can be fit by
\begin{equation}
\label{eq:kesi_vs_parameters}
  \xi=53.2\frac{t_z^2}{W^2}C+0.363.
\end{equation}
The above equation enables us to estimate the localization length under various parameters.
For example, using the parameters for the quasi-1D metal in Sec.~\ref{subsubsec:1D metal},  the calculated localization length is $\xi=106$.
The localization length is about five times the size of $L_z$, indicating the system is indeed in the metallic regime.
We note that special care is required when applying Eq.~(\ref{eq:kesi_vs_parameters}) to systems with large $t_z$ or $W$.
In these cases, the (mobility) gaps may already close, thus a well-defined surface state does not exist.

\section{Summary and Discussion}
\label{sec:conclusions}
In summary, we have systematically investigated the longitudinal conductance of a disordered 3D quantum Hall system within a tight-binding lattice model using numerical Thouless number calculations.
In particular, we find by stacking the 2D QHE layers along the $z$ direction, for small interlayer hoppings, the peak conductance of lowest subband in the $x$ direction scales linearly with the number of layers, while in the $z$ direction, it is inversely proportional to the number of layers.
We confirm that the mobility edges of the bulk are the same along the $x$ and $z$ directions for this anisotropic system.
For extended states, the longitudinal conductance along the magnetic field behaves like a quasi-1D normal metal; perpendicular to the magnetic field, the longitudinal conductance is controlled by layered conducting states stacked coherently.
Inside the quantum Hall gap, we demonstrate the crossover of the 2D chiral surface states between the quasi-1D metal and insulator regimes by modifying the interlayer hopping strength $t_z$ and the disorder strength $W$.
In real experiments, these can be achieved by uniaxial stress and disorder doping, respectively.
The typical behaviors of the Thouless conductance and the wave functions of the surface states in these two regimes are presented.
In order to predict the regime of the surface states for arbitrary parameters, we determine an explicit relationship between the localization length of surface states and the microscopic parameters of the model, which should be useful in detecting and controlling the chiral surface states in the experiments. 

Besides the very recent work of Ref.~\cite{Liu19}, the only experimental realization of the 2D chiral surface states is the engineered multilayer quantum well system~\cite{Druist98}.
However, since the sample sizes are much larger than the phase coherence lengths in the experiment, only an incoherent 2D chiral metal has been studied~\cite{Druist98,Cho97,Druist99}.
So far, the three regimes of phase-coherent transport have not been investigated in the experiments.
Recently discovered anisotropic layered 3D materials thus offer a unique opportunity to study both the 3D QHE and its novel 2D chiral surface states.
Samples with dimensions smaller than the phase coherence lengths are highly desired in the future.
We mention that for such small samples, in addition to transport measurement, real-space probe techniques such as scanning tunneling microscopy can also be a useful tool to detect the metal-insulator crossover of the surface states~\cite{Haude01,Morgenstern01}, as demonstrated in Figs.~\ref{fig:1D_metal} and \ref{fig:1D_insulator} in the paper.

\section{Acknowledgements}
\label{sec:acknowledgements}
The work at Zhejiang University was supported by the National Natural Science Foundation of China through Grant No. 11674282 and the Strategic Priority Research Program of Chinese Academy of Sciences through Grant No. XDB28000000. KY's work was supported by the National Science Foundation Grant No. DMR-1932796, and performed at the National High Magnetic Field Laboratory, which is supported by National Science Foundation Cooperative Agreement No. DMR-1644779, and the State of Florida.

\bibliography{reference}
\end{document}